\documentclass[journal=jacsat,manuscript=article]{achemso}

\usepackage{chemformula} 
\usepackage[T1]{fontenc} 

\author{Chuanwen Zhao}
\affiliation{MOE Key Laboratory of Fundamental Physical Quantities Measurement \& Hubei Key Laboratory of Gravitation and Quantum Physics, PGMF and School of Physics, Huazhong University of Science and Technology, Wuhan 430074, China}

\author{Xin Yi}
\affiliation{MOE Key Laboratory of Fundamental Physical Quantities Measurement \& Hubei Key Laboratory of Gravitation and Quantum Physics, PGMF and School of Physics, Huazhong University of Science and Technology, Wuhan 430074, China}
 
\author{Qiao Chen}
\affiliation{MOE Key Laboratory of Fundamental Physical Quantities Measurement \& Hubei Key Laboratory of Gravitation and Quantum Physics, PGMF and School of Physics, Huazhong University of Science and Technology, Wuhan 430074, China}

\author{Chengyu Yan}
 \email{chengyu_yan@hust.edu.cn}
\affiliation{MOE Key Laboratory of Fundamental Physical Quantities Measurement \& Hubei Key Laboratory of Gravitation and Quantum Physics, PGMF and School of Physics, Huazhong University of Science and Technology, Wuhan 430074, China}
\affiliation{Institute for Quantum Science and Engineering,
Huazhong University of Science and Technology, Wuhan 430074, China}

\author{Shun Wang}
 \email{shun@hust.edu.cn}
\affiliation{MOE Key Laboratory of Fundamental Physical Quantities Measurement \& Hubei Key Laboratory of Gravitation and Quantum Physics, PGMF and School of Physics, Huazhong University of Science and Technology, Wuhan 430074, China}
\affiliation{Institute for Quantum Science and Engineering,
Huazhong University of Science and Technology, Wuhan 430074, China}

\title[An \textsf{achemso} demo]
  {Josephson Effect in NbS$_{2}$ van der Waals Junctions}


\begin{document}

\begin{abstract}
Van der Waals (vdW) Josephson junctions can possibly accelerate the development of advanced superconducting device that utilizes the unique properties of two-dimensional (2D) transition metal dichalcogenide (TMD) superconductors such as spin-orbit coupling, spin-valley locking. Here, we fabricate vertically stacked NbS$_{2}$/NbS$_{2}$ Josephson junctions using a modified all-dry transfer technique and characterize the device performance via systematic low-temperature transport measurements. The experimental results show that the superconducting transition temperature of the NbS$_{2}$/NbS$_{2}$ Josephson junction is 5.84 K, and the critical current density reaches 3975 A/cm$^{2}$ at 2K. Moreover, we extract a superconducting energy gap $\Delta=0.58$ meV, which is considerably smaller than that expected from the single band s-wave Bardeen-Cooper-Schrieffer (BCS) model ($\Delta=0.89$ meV). 

\end{abstract}

\newpage

\subsection{TOC GRAPHIC}
\begin{figure}[ht]
\includegraphics[scale=1]{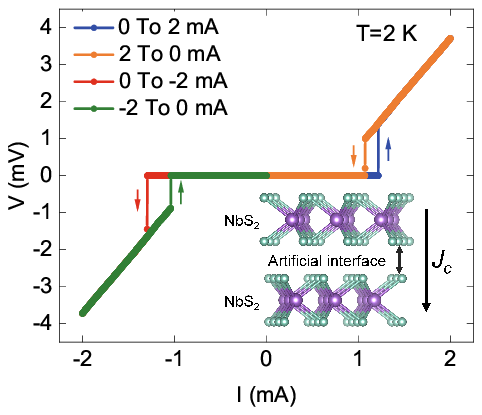}
\label{TOC GRAPHIC}
\end{figure}

The Josephson junction is one of the basis of quantum metrology, quantum communication, and quantum computing. The Josephson junctions consisted of superconductor/oxide insulator/superconductor, such as Al/Al$_{2}$O$_{3}$/Al, are the mainstream scheme which leads to some of the most celebrated breakthrough in the fields such as the demonstration of quantum supremacy\cite{arute2019quantum}. However, the in-homogeneity of the insulating layer and defect residing within the layer plague the device performance\cite{zeng2015direct,mcdermott2009materials}. For instance, these imperfections can result in noticeable decoherence in a superconducting qubit and Josephson microwave source\cite{kim2021enhanced,yan2021low}. It is therefore necessary to explore alternative platforms for realizing Josephson junction which can possibly overcome the aforementioned shortcomings. In this regard, the layered transition metal dichalcogenides (TMD), such as NbSe$_{2}$, has been one of the leading platforms, and stimulates the developments of Josephson junctions consisting of two superconducting TMD flakes, where misalignment and misorientation cause the difference in the order parameter for the two flakes and give rise to the Josephson relation (the conceptual illustration of vdW Josephson junction is shown in Figure S1)\cite{yabuki2016supercurrent,farrar2021superconducting}.

However, a comprehensive understanding of TMD-based Josephson junction is still lacking. One of the reasons is that for the superconducting TMDs, their superconductivity coexists with charge-density wave (CDW) states\cite{xi2015strongly}. For instance, O. Naaman et al. found that in Pb/I/NbSe$_{2}$ junction, the superconducting energy gap of NbSe$_{2}$ is smaller than expected, which could be attributed to the interplay between the superconductivity and the coexisting CDW in this material\cite{naaman2003josephson}. This can potentially complicate the implementation of all TMD-based superconducting circuit. Fortunately, 2H-NbS$_{2}$ shares the same crystal structure with NbSe$_{2}$ and it is the only superconducting TMD material with no CDW order observed experimentally\cite{naito1982electrical}. Understanding the behavior of NbS$_{2}$ Josephson junction may provide a universal insight to the key properties of other TMD-based Josephson junctions. However, high-quality NbS$_{2}$ Josephson junctions are less studied\cite{majumdar2020interplay}. The reasons are twofold: First, the exfoliated NbS$_{2}$ thin flakes are not stable and suffers rapid oxidization in ambient\cite{yan2019thickness}. Second, the interlayer contaminants within vdW junctions may lead to the degradation of device quality\cite{schwartz2019chemical}.

In this letter, we report the observation of Josephson effect in high quality NbS$_{2}$/NbS$_{2}$ vdW junctions for the first time. The junction is obtained by a modified all-dry transfer technique under inert environmental conditions. The critical current density is as high as 3975 A/cm$^{2}$ at 2 K, which may meet the need for applications including high output on-chip microwave source\cite{yan2021low}, high gain Josephson amplifier\cite{narla2014wireless}. The extracted superconducting energy gap is smaller than the value predicted by the BCS theory. Our results call for more detailed investigation on the superconducting mechanism in NbS$_{2}$ and other TMD materials in general, which will promote the development of TMD superconducting circuits.

\begin{figure}[ht]
\includegraphics[scale=0.9]{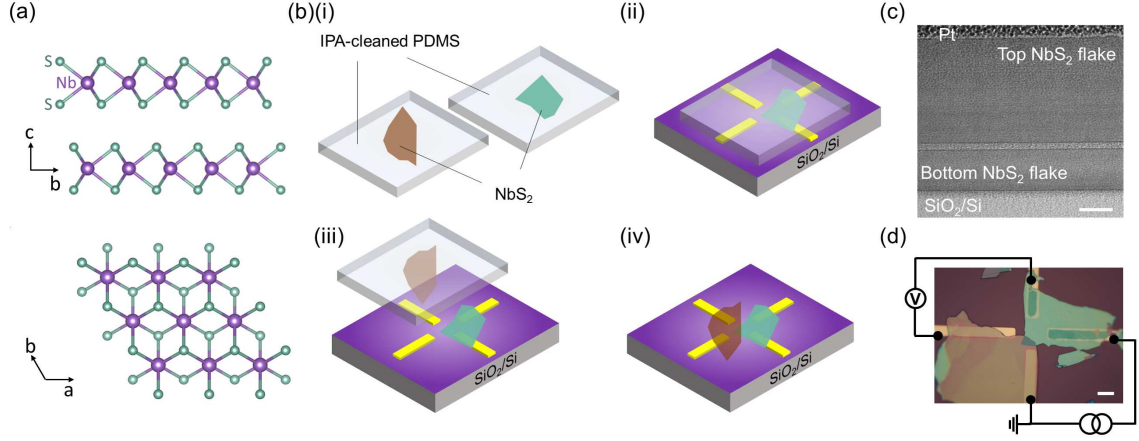}
\caption{vdW Josephson junction based on 2H-NbS$_{2}$. (a) Schematic crystal structure of 2H-NbS$_{2}$ (Nb, purple and S, green). The top (upper) and side (bottom) views show the layered structures. (b) Device fabrication process of vertical NbS$_{2}$/NbS$_{2}$ Josephson junction. (i) Exfoliation of NbS$_{2}$ flakes on IPA-cleaned PDMS stamps. (ii-iv) Transfer two pieces of NbS$_{2}$ flake onto the silicon substrate with Au/Ti contacts (45 nm of Au and 5 nm of Ti) with PDMS stamps. (c) Cross-sectional TEM analysis of a junction device over the entire region cut out by focused ion beam. And the device composed of top and bottom NbS$_{2}$ flakes, with a thickness of 60.9 nm and 24.3 nm, respectively. The scale bar labels 20 nm. (d) Optical micrograph of the mechanically assembled NbS$_{2}$/NbS$_{2}$ junction with four-terminal measurement conﬁguration. The junction area is 31.4 $\mu\mathrm{m}^{2}$. Current is applied between the two flakes and voltage drop is recorded. The scale bar labels 10 $\mu\mathrm{m}$.}
\label{fgr:1}
\end{figure}

High-quality 2H-NbS$_{2}$ single crystals were grown by chemical vapor transport using iodine as a transport agent and had a typical T$_{c}$ of $\sim$6.13 K\cite{lian2017effect}. Figure 1(a) shows the top and side view of the layered structure of 2H-NbS$_{2}$, respectively.  Vertical NbS$_{2}$/NbS$_{2}$ junction devices are fabricated by a modified dry-transfer technique in a Nitrogen-filled glove box to avoid contamination and oxidization. Water and oxygen are evacuated below 1 ppm for the most optimized device quality. The detailed device fabrication processes are illustrated in Figure 1(b). First, the bulk crystals are mechanically exfoliated and placed on top of silicone elastomer polydimethylsiloxane (PDMS) substrates using adhesive tape. It is necessary to emphasize that the PDMS substrates are cleaned with isopropanol alcohol (IPA) before transferring the NbS$_{2}$ flakes onto it, ensuring a clean interface\cite{schwartz2019chemical}. Then flakes of uniform surface and desired geometry are chosen and sequentially stacked onto a SiO$_{2}$/Si substrate with pre-patterned Au/Ti electrodes. To remove any resist residues in the contact areas before transfer, the wafers were cleaned with acetone, IPA, deionized water and oxygen plasma at 300 W for 5 min. To avoid the effect of heating on the surface degradation of the NbS$_{2}$ flakes, the entire device fabrication process was performed at room temperature. The thickness of all the NbS$_{2}$ flakes used for the devices in this work are between 20 and 60 nm, to ensure negligible suppression T$_{c}$ (Figure S2), because the T$_{c}$ of NbS$_{2}$ is known to decrease in ﬂakes \textless10 nm\cite{yan2019thickness}. Figure 1(c) shows a transmission electron microscope (TEM) image of the vertical cross-section of the junction, an artificial interface (equivalent to the insulating layer) with a thickness of $\sim$3 nm resides between the top and bottom NbS$_{2}$ flakes. Some non-pristine interface layers and artificial interface are likely to be NbO$_{x}$, which may be caused by the oxidization processing during taking the TEM images. To obtain the TEM images, the device was cleaved by focused ion beam and exposed in the ambient environment, and the oxidization came into play. The air-sensitivity of NbS$_{2}$, especially compared to NbSe$_{2}$, renders the difficulty in obtaining a high-quality junction. However, we would like to stress that the oxidized interface does not necessarily occur in the measured sample where it is kept in high vacuum condition. Figure 1(d) is a schematic illustration of the junction structure with a four-terminal measurement configuration and the typical junction areas span from 7.8 to 50 $\mu\mathrm{m}^{2}$.

\begin{figure}
\includegraphics[scale=0.9]{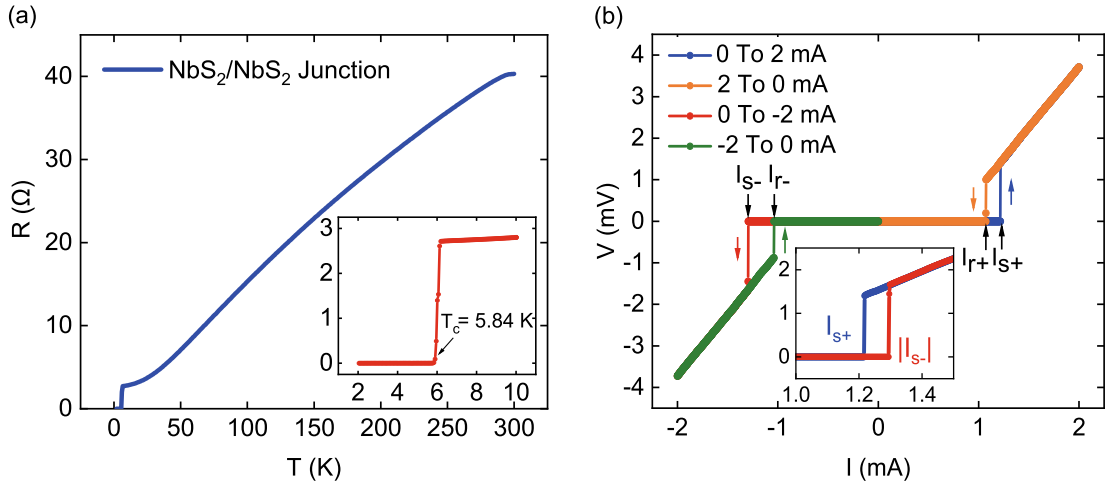}
\caption{ Characterization of a NbS$_{2}$ vdW Josephson junction. (a) Temperature-dependent resistance of the junction device. The inset shows a zoomed-in view of the superconducting transition. The inset shows a zoomed-in view of the superconducting transition. The critical temperature T$_{c}$ is 5.84 K. (b) Current-voltage (I-V) characteristic curve measured across the junction at temperature $T=2$ K. Current bias is swept both forward and backward. The I$_{s}$ and I$_{r}$ in the positive and negative current regime are labelled I$_{s+}$, I$_{r+}$, I$_{s-}$ and I$_{r-}$, respectively. The inset presents the absolute value I$_{s+}$ and I$_{s-}$ of the I-V curve of JJ1.}
\label{fgr:2}
\end{figure}

We first measured the temperature dependence of  four-terminal resistance R(T) of a Josephson junction (JJ1) in the presence of a 10 $\mu\mathrm{A}$ d.c. excitation current at zero magnetic field and observed the superconducting transition in Figure 2(a). The critical temperature T$_{c}$ of this device is determined to be 5.84 K. Note that T$_{c}$ is defined as the temperature where resistance drops to 10$\%$ of the normal-state resistance. The transition window is as narrow as 0.2 K, without any additional kinks as shown in the inset of Figure 2(a). The clean transition highlights the cleanness of the junction area thanks to our specifically designed transfer protocol.

Next, the current-voltage (I-V) curve of JJ1 measured at 2 K is plotted in Figure 2(b), which displays a pronounced hysteretic feature, a signature of an under-damped Josephson junction. We checked that similar hysteresis behavior can be obtained at different current sweeping rates, which rules out the possibility that the observed hysteresis arises from Joule heating. This suggests that the vdW interface decouples the superconducting order parameters and creates a weak link that allows the flow of a Josephson supercurrent. We observed the same hysteretic feature in several other Josephson junctions (Figure S3). Besides, the switching current I$_{s}$ is about 1.26 mA in Figure 2(b), which corresponds to a current density of 3975 A/cm$^{2}$. The large switching current and current density are indication of strong Josephson coupling, which would be beneficial for applications including superconducting quantum qubit\cite{clarke2008superconducting} and Josephson microwave source\cite{yan2021low}. The Stewart-McCumber parameter\cite{tinkham2004introduction} of the device is estimated to be $ \beta_{\mathrm{c}} \approx\left(\frac{4 I_{\mathrm{s}}}{\pi I_{\mathrm{r}}}\right)^{2} \approx 2.31$ (where $ I_{r}=1.06 $ mA is the re-trapping current) and the junction capacitance calculated by equation $ \beta_{c}=2 e I_{s} R_{\mathrm{N}}^{2} C / \hbar $ (where $ R_{\mathrm{N}} $=2.90 $\Omega$ is normal state resistance of the Josephson junction extracted from I-V curve) is 0.23 $\mu\mathrm{F} \mathrm{cm}^{-2}$. The characteristic parameters obtained from other junctions are presented in Table S1. It is interesting to point out that both the switching current and re-trapping current can be asymmetric, and the asymmetry lager than the fluctuation of I$_{s}$ (the results of other junctions are shown in Figure S3 and the switching current statistics of JJ3 can be found in Figure S4). This observation might be caused by the superconducting diode effect (SDE)\cite{ando2020observation,narita2022field} or the self-field effect arising from the superconducting electrodes\cite{barone1975current,schwidtal1969barrier,barone1982physics}. However, for our NbS$_{2}$ Josephson junctions, the asymmetry is insensitive to the out-of-plane magnetic field, which contrary to the predicted behavior of SDE arising from Ising spin-orbital coupling (the dominate type spin-orbital coupling in NbS$_{2}$). As a result, we believe that the nonreciprocal supercurrent is not caused by SDE. Instead, we attribute the observation to the self-field effect. Interestingly the asymmetric switching current has not been reported in the more intensively studied NbSe$_{2}$\cite{yabuki2016supercurrent}.

\begin{figure}[ht]
\includegraphics[scale=1]{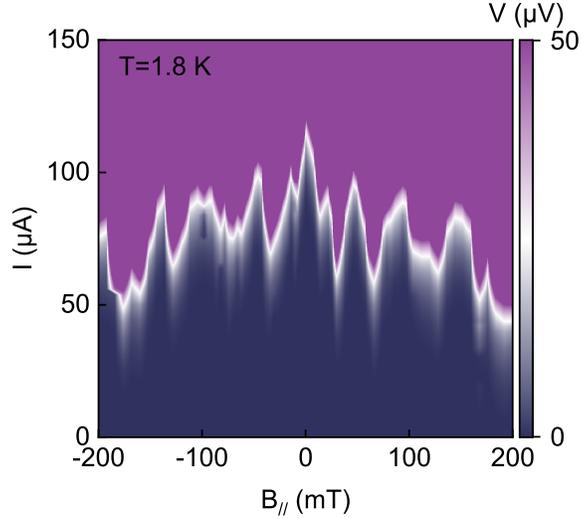}
\caption{ Contour plot of voltage as a function of bias current and external parallel field measured at T=1.8 K in JJ4 ($\sim$7.8 $\mu\mathrm{m}^{2}$, encapsulated by h-BN).}
\label{fgr:3}
\end{figure}

We further examined the field effect of switching current for our Josephson junctions by applying an in-plane magnetic field. Figure 3 shows the contour plot of voltage as a function of bias current and external parallel field measured at a temperature of 1.8 K in JJ4 when current is swept from 0 to 150 $\mu\mathrm{A}$. A periodic modulation resembling the expected Fraunhofer pattern is observed. The minima of I$_{s}$(B) do not reach zero, which may be due to the flux focusing\cite{gu1979properties,rosenthal1991flux,island2016thickness,paajaste2015pb} and the uneven interface in the vicinity of the electrodes (see Figure S5 of the Supporting information for the cross-sectional TEM image), like the results reported in MgB$_{2}$ break junctions\cite{tao2003josephson} and high-T$_{c}$ superconductor grain boundary junction.\cite{rosenthal1991flux,nicoletti1997electrical} The asymmetric data is most likely caused by the bias current induced magnetic field.\cite{kim2017strong,barone1982physics} Additionally, the London penetration depth ($\lambda_{L}$) of NbS$_{2}$ can be obtained from $ \Phi_{0}=B_{0} W(d+2 \lambda_{L}) $, where $ \Phi_{0} $ is the magnetic flux quantum, $B_{0}$ is the period of the oscillation in the Fraunhofer pattern, $ W $ is width of the junction perpendicular to the field direction, and d is the interface thickness. From the Fraunhofer pattern, the modulation period is averaged to be around 45 mT after ignoring some small oscillations and with $ W\approx2.2\mu\mathrm{m}$, and $ d\approx3 nm$, which corresponds to a $\lambda_{L}$ is calculated to be 9 nm, an order smaller than $\lambda$ in NbS$_{2}$ bulk\cite{leroux2012anisotropy}. We did not observe a well-defined periodic modulation in junctions with larger overlap area, such as JJ1, mainly due to the in-homogeneity in the junction interface.

\begin{figure}
\includegraphics[scale=0.9]{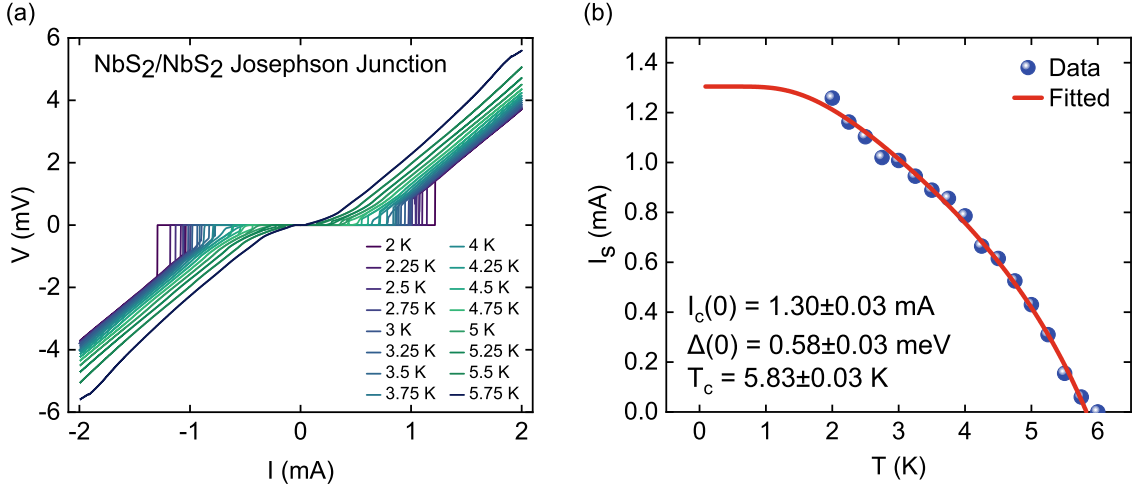}
\caption{(a) Change in the current-voltage (I-V) characteristics for temperatures between 2 K and 6 K. (b) Temperature dependence of the switching current I$_{s}$. The solid line is calculated from AB theory.}
\label{fgr:4}
\end{figure}

The superconducting mechanism in TMD material is under debate, here we provide further insight to this topic by measuring the superconducting energy gap of NbS$_{2}$. We measured the I-V curves of the JJ1 at temperatures between 2 K and 6 K (see the curves in Figure 4(a)). Figure 4(b) shows the temperature dependence of I$_{s}$, in which the solid curve is calculated by the Ambegaokar-Baratoff (AB) theory\cite{ambegaokar1963tunneling}, in which I$_{s}$(T) is expressed as 

\begin{equation}
 \frac{I_{\mathrm{s}}(T)}{I_{\mathrm{c}}(0)}=\frac{\Delta(T)}{\Delta(0)} \tanh \left[\frac{\Delta(T)}{2 k_{\mathrm{B}}T}\right]
\end{equation}

Where $\Delta(T)=\Delta(0)\tanh\left(2.2\sqrt{T_{c}/T-1}\right)$. Here $\Delta(0)$ and T$_{c}$ are fitting parameters. T$_{c}$ is found to be 5.83 K and in good agreement with the that of bulk NbS$_{2}$ ($\sim6.13$ K), indicating no degradation of superconductivity at the vdW Josephson junction. The fitted superconducting energy gap $\Delta(0)=0.58$ meV is much smaller than the value of the standard BCS ratio $1.76k_B T_c=0.89$ meV. Previous studies point out that NbS$_{2}$ exhibits two superconducting energy gaps\cite{majumdar2020interplay,guillamon2008superconducting,kavcmarvcik2010specific,kavcmarvcik2010studies}. It is particularly interesting to note that the  small gap falls below the BCS value, as confirmed by different measurement techniques including Andreev reflection spectroscopy\cite{majumdar2020interplay}, scanning tunneling microscopy and spectroscopy\cite{guillamon2008superconducting}, speciﬁc heat measurements\cite{kavcmarvcik2010specific}, ac-calorimetry measurements\cite{kavcmarvcik2010studies}. The reported small gap ranges from 0.53 meV to 0.56 meV, in good agreement with our results. In NbSe$_{2}$, it is speculated that the discrepancy between the measured gap and the BCS value is originated from the interaction between CDW and superconductivity\cite{naaman2003josephson}. Hence, our work in conjunction with previous studies suggest that even in the absence of CDW, the standard BCS theory cannot capture the superconducting behavior in NbS$_{2}$ and TMD superconductor in general may arise from mechanisms other than CDW. 

In conclusion, we successfully fabricated vdW-stacked vertical NbS$_{2}$/NbS$_{2}$ Josephson junctions by using a modified all-dry transfer technique. We confirmed the stacked structure indeed exhibits a pronounced Josephson effect via systemically transport measurements. The Josephson coupling was found to be strong, manifested as the large critical current, rendering the platform suitable for single flux quantum circuit. We also extract  superconducting gap of 0.58 meV which is considerably smaller than the BCS prediction, this may give hints on the long-standing arguments on superconducting mechanism in NbS$_{2}$ and other TMD in general.

\begin{acknowledgement}
We acknowledge the support from the National Natural Science Foundation of China (12074134) and the Analysis and Testing Center of Huazhong University of Science and Technology.
\end{acknowledgement}

\begin{suppinfo}

The following files are available free of charge.

Concept of a vdW Josephson junction; RT curve of NbS$_{2}$ Bulk and flake device; I-V curves of additional NbS$_{2}$/NbS$_{2}$ junctions; Switching current statistics of JJ3; TEM image (PDF)

\end{suppinfo}

\bibliography{ref.bib}

\end{document}